\documentclass[12pt,preprint]{aastex}
\usepackage{graphicx}
\usepackage{longtable}
\usepackage{multirow}
\usepackage{url}
\usepackage{amsmath}
\usepackage{lipsum}
\usepackage{rotating}
\usepackage{color}

\begin{document}


\title{The correlation between isotropic energy and duration of gamma-ray bursts}

\author{Z. L. Tu\altaffilmark{1} and F. Y. Wang\altaffilmark{1,2}{*}}
\affil{
$^1$ School of Astronomy and Space Science, Nanjing University, Nanjing 210093, China\\
$^2$ Key Laboratory of Modern Astronomy and Astrophysics (Nanjing University), Ministry of Education, Nanjing 210093, China\\
} \email{{*}fayinwang@nju.edu.cn}

\begin{abstract}
In this paper, we study the correlation between isotropic energy and
duration of gamma-ray bursts (GRBs) for the first time. The
correlation is found to be $T_d \propto {E_{iso}}^{0.34\pm 0.03}$
from the {\em Swift} GRB sample. After comparing with solar flares from
{\em RHESSI} and stellar superflares from {\em Kepler} satellite, we
find that the correlation of GRBs shows similar exponent with those
of solar flares and stellar superflares. Inspired by the physical
mechanism of solar flares and stellar superflares which is magnetic
reconnection, we interpret the correlation using magnetic
reconnection theory. This similarity hints
that magnetic reconnection may dominate energy releasing process of
GRBs.
\end{abstract}

\section{Introduction}
Gamma-ray bursts (GRBs) are explosive phenomena occurring at
cosmological distance \citep{2015Kumar,2015Wang,2016Zhang},
and play a vital role in multi-messenger astronomy \citep[][etc.]{2017Willingale}.
As the central engine is still mysterious, plenty of works have been
proposed to explain the burst prompt emission mechanism or central engine \citep{2017Beloborodov, 2017Dai, 2018Nagataki}, including
the internal shock model \citep{1994Rees,2011Daigne}, the
dissipative photosphere model \citep{2001Spruit,2005Rees}, the
electromagnetic model \citep{2003Lyutikov,2006Lyutikov} and the
internal-collision-induced magnetic reconnection and turbulence
model \citep{2011Zhang}. Although they have successfully interpreted
some remarkable GRBs, there are still some open questions at meanwhile.

Because there are many models for interpreting the physical
mechanism of GRBs, it is essential to statistically analyze basic
properties of GRBs. Some empirical correlations have been
found \citep{2015Wang, 2017DainottiDel, 2018DainottiAmati,
2018Dainotti}, e.g. the isotropic energy and spectral peak energy
correlation $E_{\rm iso}-E_{p}$ \citep{2002Amati,2016Wang}, the
correlation between $E_{\rm iso}$, spectral peak energy $E_{p}$ and
rest-frame break time $t_b$ \citep{2005Liang}, X-ray
luminosity $L_{X}$ and rest-frame time of plateau phase $T_{a}^{*}$
\citep{2015Dainotti}, lag-luminosity correlation $\tau_{\rm
lag}-L_{\rm iso}$ \citep{2000Norris}, and fundamental plane of GRBs
\citep{2016Dainotti,2017bDainotti}. But the correlation between
duration and isotropic energy has not been studied in literature
before.

Stellar superflares are violent energy releasing events occurring on
stellar surface. \citet{2012Maehara} statistically studied the
superflares from solar-type stars, in virtue of continuously long
periods observation of {\em Kepler}. Both long and short cadence
data from {\em Kepler} have been collected later to study
superflares in comparing with solar flares
\citep[etc.]{2013Shibayama, 2015Maehara}. To be specific,
\citet{2015Maehara} fitted the correlation between duration and
energy of stellar superflares as $\tau_{\rm flare} \propto E_{\rm
flare}^{0.39 \pm 0.03}$, which is comparable to statistical analysis
of solar flares.

In this paper, the correlation between duration and isotropic energy
of GRBs is fitted. While filling a vacancy of statistical
correlation studies of GRBs, we explore if there are some
resemblances between stellar flares and GRBs. In order to make
comparison, solar flares from {\em RHESSI} and superflares of
solar-type stars from {\em Kepler} are gathered in this work.
Linear regression has been made with different kinds of data. In
order to test if our fitting results are strongly credible,
statistical methods $t$-test and $F$-test are involved. Through this
paper, we adopt cosmological parameters as $H_{0}$=67.7km/s/Mpc,
$\Omega_{m}=0.31$, $\Omega_{\Lambda}=0.69$ \citep{2016Planck}.

The paper is structured as follows. Samples of GRBs, solar flares
from {\em RHESSI}, and superflares of solar-type stars from {\em
Kepler} are presented in section \ref{sec:Data}. The results of linear
regression in log-log fields are given in section \ref{sec:mandr}.
In section \ref{sec:theory correlation}, we give a
reasonable explanation of the correlation basing on magnetic
 reconnection. Conclusions and discussion are given in section
\ref{sec:conclusions}.

\section{Data Samples}\label{sec:Data}
This section will specifically introduce the filter conditions of
data selection. Methods of calculation for different datasets will
also be presented.

\subsection{Gamma-ray bursts}\label{sec:GRBdata}
{\em Swift} has been operated to observe GRBs from 2004, and over
1,000 GRBs are detected. In this paper, we target on selecting those
GRBs with redshift measurements during January 2005 to May 2018. In
order to avoid importing system errors from different GRB surveys,
we only use GRBs from the website of {\em Swift} data table
\footnote{https://swift.gsfc.nasa.gov/archive/grb\_table/}. Besides
of redshift, duration time $T_{90}$ and fluence of prompt emission
$S_{\rm GRB}$ with $90\%$ error can also be obtained from web site.
Those GRBs showing absence of duration time, error of fluence are
all excluded from the database. Then, we use the redshift
measurements for GRBs from Jochen Greiner's
website\footnote{http://www.mpe.mpg.de/{$\sim$}jcg/grbgen.html}.
There are total 386 GRBs for further study.

Because GRBs occur at cosmological distances, the time should
be transferred to rest frame. The duration can be written as
\begin{equation}\label{equ:GRBduration}
    T_{\rm GRB,duration}=\frac{T_{90}}{1+z}.
\end{equation}
The isotropic energy is
\begin{equation}
\label{equ:GRBenergy}
E_{\rm GRB,iso}=\frac{4\pi D_{\rm L}^{2} S_{\rm GRB}}{1+z},
\end{equation}
where $D_{\rm L}$ is luminosity distance which relates to cosmological parameters. $S_{\rm GRB}$ represents fluence.

\subsection{Solar flares}\label{sec:Solarflaredata}
Solar flares have been studied for lengthy period from the first
observation in 1859 \citep{1859Carrington}. {\em RHESSI} spacecraft
is manipulated successfully over 16 years. More than 120,000 solar
flares are observed from 2002 to 2018, which are listed
online\footnote{https://hesperia.gsfc.nasa.gov/hessidata/dbase/hessi\_flare\_list.txt}.
Bolometric energy of flare can be calculated by summing energy of
each detected photons. In order not to inaccessibly obtain energy
spectrum of each flares, we use the total counts to represent
 total energy of solar flares. The energy is proportional to the total
 counts, which
can be expressed as
\begin{equation}\label{equ:solar counts}
 E_{\rm flare} \propto C_{\rm total}.
\end{equation}
Because instrumental sensitivity gets down below $\sim5$ keV band,
the total counts are taken from 6$-$12 keV energy range
\citep{2008Christe}. The online list includes some flags marking
non-solar event (NS) and possible solar flare (PS). After
eliminating flares marked as NS or PS, 114,728 solar flares are
imported in this work. Their total counts $C_{\rm total}$ and
duration time $T_{\rm flares,duration}$ are directly obtained from
flare list.

\subsection{Superflares of solar type star}
In this paper, we also use superflares of solar-type stars.
\citet{2015Maehara} selected 23 solar-type stars with 187 white
light superflares from 18 quarters of {\em Kepler} short-cadence
data. We obtain the properties of these superlares, including energy
of flares $E_{\rm superflares}$ and duration $T_{\rm superflares}$
\citep{2015Maehara}. To be specific, the duration $T_{\rm
superflares}$ is derived from $e$-folding decay time.
\\
\section{Methods and results}\label{sec:mandr}
\subsection{Linear regression}
We use the duration and energy from different datasets
to constrain the power-law correlation
\begin{equation}
T=10^{b} \times E^{k},
\end{equation}
where energy $E$ is substituted as $C_{\rm total}$ for solar flares,
and $T$ represents duration of different data. We take $y=\log_{10}T$ and $x=\log_{10}E$ in log-log fields. Then this correlation can be derived as
\begin{equation}
y=kx+b,
\end{equation}
where $k$ and $b$ are fitted from linear regression in this work.

We use maximum likelihood method to perform linear regression. The
general likelihood can be written as \citep{2005D'Agostini}
\begin{equation}
\mathcal{L}\propto\prod\limits_{i}\frac{1}{\sqrt{\sigma_{v}^{2}+\sigma_{y_{i}}^{2}+
k^{2}\sigma_{x_{i}}^2}}\exp
\left[-\frac{\left(y_{i}-kx_{i}-b\right)}{2\left(
\sigma_{v}^{2}+\sigma_{y_{i}}^{2}+k^{2}\sigma_{x_{i}}^2\right)}\right],
\end{equation}
where $\sigma_{v}$ is extra variability. $\sigma_{x_{i}}$ and $
\sigma_{y_{i}}$ are variants taken from observations. We take
$\sigma_{x_{i}}=0$ and $ \sigma_{y_{i}}=0$ for solar flares and
superflares of solar-type stars, due to the errors are not included
in database.

\subsection{Testing significance of regression}
In order to test significance of regression, $t$-tests and $F$-tests
methods are imported in this work.
In a short words, these methods are used to quantitatively test
whether the slope of linear regression can reject null slope
hypothesis, and describe compactness between the slope of
correlation and data. The $t_{\rm value}$ can be written as
\citep{2012Montgomery}
\begin{equation}
t_{\rm value }=\frac { \left(k-k_{ 0 } \right) \sqrt { \left( n-2 \right)
\sum \limits_{i=1}^{n}\left( x_{i}-{ \bar { x }  } \right) ^{ 2 } }  }{ \sqrt { \sum \limits_{i=1}^{n}\left( y_{i}-kx_{i}-b \right) ^{ 2 } }   } ,
\end{equation}
where $k$ and $b$ are results of regression. $x_{i}$ and $y_{i}$ are
properties from observational data. $n$ is the number of data
points. And $\bar {x}$ gives the mean value of $x_{i}$. The null
hypothesis means $k_{0}=0$. If
\begin{equation}
\left|t_{\rm value}\right| >t_{\alpha/2,n-2},
\end{equation}
the null hypothesis is rejected at upper percentage point, where
$t_{\alpha/2,n-2}$ represents the rejection regions of $t$
distribution. Here we take $\alpha = 5\%$.

$F$-test is also imported in this work as replenishment of $t$-test.
The $F_{\rm value }$ can be written as \citep{2012Montgomery}
\begin{equation}
F_{\rm value} = \frac{\left(n-2\right)\sum
\limits _{i=1} ^{n}\left( kx_{i}+b-{\bar y}\right)^{2}}{\sum \limits_{  i=1}^{ n } \left( y_{i}-kx_{i}-b \right)^{2} },
\end{equation}
where $\bar {y}$ gives mean of $y_{i}$. Refer to $t$-test, we take
the $F_{\alpha,1,n-2}$ of $F$ distribution as rejection regions. If
\begin{equation}
F_{\rm value} >F_{\alpha,1,n-2},
\end{equation}
the null hypothesis is rejected. $\alpha = 5\%$ is also applied for $F$-test.

\subsection{Results}
Results of regression and statistical variances of different
datasets are presented in Table \ref{tab:result}. For each kind of
data, power-law relation between duration and releasing energy are
fitted.

\begin{figure*}
\centering
\includegraphics[width=0.6\linewidth]{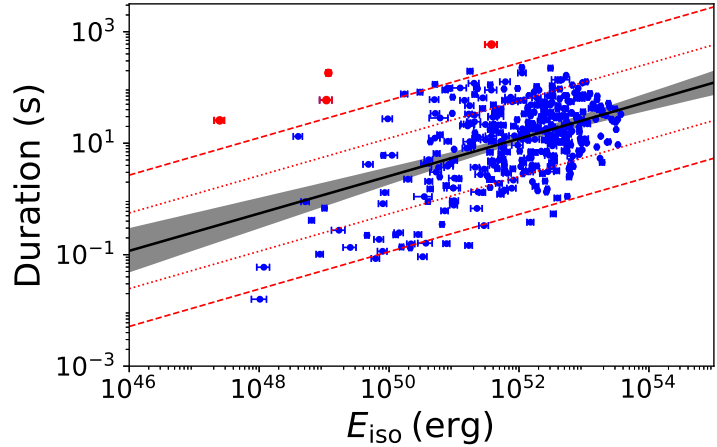}
\caption{Linear fitting of GRBs. Black solid line represents the
    fitting result in log-log field. Red dotted and dashed lines represent $1\sigma_{v}$
      and $2\sigma_{v}$ regions of extra variabilities, and gray area stands for 95\% confidence
       interval of fitting uncertainties.
Four red points located far outside of $2\sigma_{v}$ region
are GRB 111005A, GRB 080517, GRB 101225A and GRB 171205A.}
\label{fig:grbline}
\end{figure*}

After applying linear regression in log-log fields, we get the relation of GRBs as
\begin{equation}
\label{res.GRB}
T_{\rm GRB,duration} \propto {E_{\rm GRB,iso}}^{0.34\pm 0.032}.
\end{equation}
Figure \ref{fig:grbline} gives result of linear regression.
The correlation coefficient is $r=0.47$, and
$\sigma_{v}=0.66\pm0.024$. The result of $t$-test is $\left| t_{\rm
value}\right|\approx 10.32$, which is bigger than the rejection
region at $t_{2.5\%,308-2}=1.97$. $F$-test gives $F_{\rm
value}\approx 106.58$, which is greater than
$F_{5\%,1,308-2}=3.87$. This results strengthen our believe that
the duration is correlated with isotropic energy. In Figure
\ref{fig:grbline}, four red points are located outside of
$2\sigma_{v}$ area, which are GRB 111005A, GRB 080517, GRB
101225A and GRB 171205A. They are extraordinary GRBs showing low
luminosity or ultra-long duration
\citep{2014Levan,2015Stanway,2017Dado,2018MichalowskI}.

It must be noticed that the selection biases may be
important, which is out the scope of
 this work. \citet{2013Kocevski} used simulated GRBs
    to propose that duration of GRBs may not be dilated by cosmological
     expansion but decreased by detectors, due to the diminishing ratio of
      signal-to-noise. At hight redshift, only the brightest GRBs can be
       detected. Some works concluded that
        duration of GRB is mainly affected by cosmological dilation
         \citep[e.g.][]{2013Zhang,2014Littlejohns}. Recently,
         \citet{2018Lloyd} found an anti-correlation between source frame
          durations and redshifts of radio-loud GRBs. So in the future, it is
           required to apply the \citet{1992Efron} method which has been
            broadly used \citep[e.g.][]{1999Lloyd,2013Dainotti,2015Dainotti,2015Yu,2017aDainotti,2018Zhang}
             to reveal the nature of $T_{\rm GRB,duration}-{E_{\rm GRB,iso}}$
              correlation.

\begin{figure*}
\centering
\includegraphics[width=0.6\linewidth]{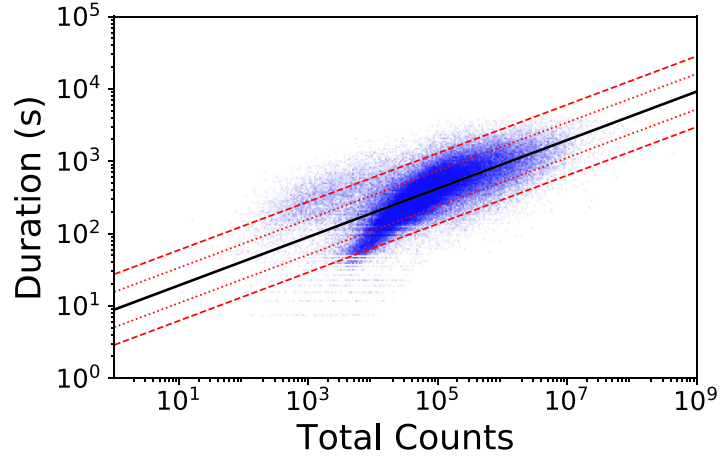}
\caption{Linear fitting of solar flares. Black solid line represents
the result of fitting in log-log field. Red dotted and dashed lines represent $1\sigma_{v}$ and $2\sigma_{v}$ regions of extra variabilities, and gray area stands for 95\%
     confidence interval of fitting uncertainties. Because error of this fitting result is tiny, black
  solid line, and gray area are overlapped.}
\label{fig:solarfit}
\end{figure*}

Linear regression of solar flares gives
\begin{equation}
\label{res.solar}
T_{\rm solar,duration} \propto {C_{\rm total}}^{0.33\pm 0.001}.
\end{equation}
This correlation is compatible with the result of solar flares
\citep{2002Veronig}. Note that the fitting errors are very tiny.
Therefore, 95\% confidence regions of fitting uncertainties, and fitting line are overlapped in Figure \ref{fig:solarfit}.
The correlation coefficient is $r=0.72$, which indicates that the
dependency between fitting line and data is moderate. $t$-test and
$F$-test also prove this dependency is obvious, where $t_{\rm
value}\approx 350.82$ and $F_{\rm value}\approx 123076.52$ are much
larger than $t_{ 2.5\%,114728-2}=1.96$ and
$F_{5\%,1,114728-2}=3.84$.

\begin{figure*}
\centering
\includegraphics[width=0.6\linewidth]{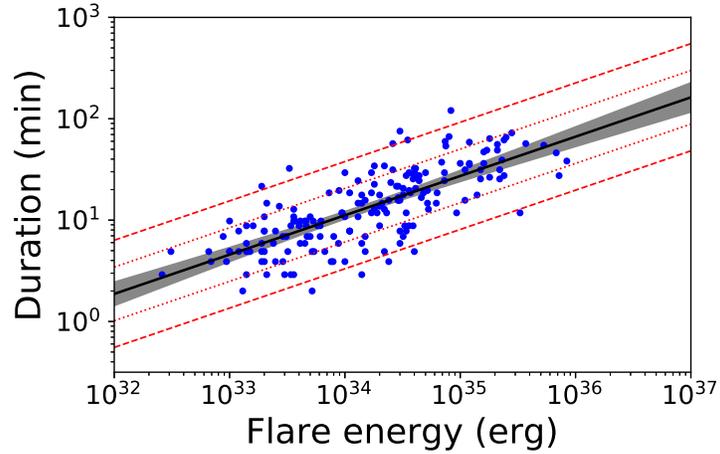}
\caption{Linear fitting of stellar superflares. Black solid line
represents the best fitting in log-log field. Red dotted and dashed lines represent
     $1\sigma_{v}$ and $2\sigma_{v}$ regions of extra variabilities, and gray area stands for
      95\%
    confidence interval of fitting uncertainties.}   \label{fig:sflare}
\end{figure*}

We use properties of superflares from \citet{2015Maehara}.
In contrast to their work,
we obtain an identical correlation in
Figure \ref{fig:sflare}. Note that unlike what we get above for
other datasets, the duration here is in unit of minutes.
This is completely unrelated to the slope of linear regression.
Linear fitting gives
\begin{equation}
\label{res.star}
T_{\rm superflares} \propto {E_{\rm superflares}}^{0.39\pm 0.025}.
\end{equation}
Here, the correlation coefficient is $r=0.75$. $t_{\rm value}\approx
15.48$, and $F_{\rm value}\approx 239.52$ are much larger than
$t_{2.5\%,187-2}=1.97$, and $F_{5\%,1,187-2}=3.89$ respectively. So
this linear correlation is strongly subsistent.

\section{Correlation between duration and energy}
\label{sec:theory correlation} In the above section, we ¡¥nd that
the slopes of the correlations for GRBs, solar ¡ãares and
super¡ãares are similar, which indicates the physical mechanism for
these phenomena is similar. We try to explain the slope of GRBs
using magnetic reconnection theory.

We make the first assumption that GRBs release magnetic energy stored in
central engine. This assumption is similar to solar flares, which release
magnetic energy stored near sun spots. The relationship between
releasing energy and magnetic energy can be written in as
\begin{equation}
\label{equ:E'}
E_{GRB} \sim fE_{mag}\propto fB^{2}L^{3}\sim fB^{2}V_{mag},
\end{equation}
where $f$ represents the fraction of energy released by magnetic
dissipation. $L$ corresponds to the typical length of
magnetic reconnection scale, and $L^{3}$ represents volume $V_{mag}$,
where magnetic energy is stored.

Moreover, with a view of that the flare energy is mainly released through
magnetic reconnection, duration of energy releasing can be
comparable with magnetic reconnection time. And this relation can be
expressed as
\begin{equation}
\label{equ:magnetic energy}
T_{\rm duration}\sim \tau_{\rm rec} \sim \frac{\tau_{\rm A}}{M_{\rm A}} \sim \frac{L}{v_{\rm A}M_{\rm A}},
\end{equation}
where $\tau_{\rm A}=L/v_{\rm A}$ represents time of plasma traveling
with ${\rm Alfv\acute{e}n}$ speed. ${\rm Alfv\acute{e}n}$-Mach
number $M_{\rm A}$ stands for reconnection rate, which can be
treated as a constant. For GRBs, ${\rm Alfv\acute{e}n}$ speed may
close to the speed of light, namely, $v_{\rm A}\sim c$
\citep{1975Jackson,1999Lazarian}.
Naturally, the relation between duration and releasing
energy for one GRB can be expressed as
\begin{equation}
\label{equ:theoretical function}
T_{\rm duration} \propto E^{\frac{1}{3}}.
\end{equation}
The correlation is comparable to what we have obtained in section
\ref{sec:mandr}.

Magnetic reconnection driving solar flares is widely accepted from
theorems and observations \citep[etc.]{1982Priest,1992Tsuneta}. We
find the correlation of GRBs as $T_{\rm GRB,duration} \propto
{E_{\rm GRB,iso}}^{0.34\pm 0.032}$,
 and the exponent $0.34\pm 0.032$ is also compatible with the exponent
 $1/3$ in eq.(\ref{equ:theoretical function}). So, our results hint
 that magnetic reconnection may also dominate energy releasing of
 GRBs during prompt emission. To be specific, our findings may support
 some remarkable works, which set magnetic reconnection as
 mechanism of powering GRBs emission
 \citep[etc.]{2011Zhang,2011Metzger,2014Zhang,2018Beniamini}.
 Interestingly, theory \citep{2006Dai} and observations
 \citep{2013Wang} also support that the magnetic reconnection also
 account for the X-ray flares of GRBs.

\section{Conclusions and discussion}
\label{sec:conclusions} In this work, we find the power law
correlation between isotropic energy and duration of GRBs for the
first time. Linear fitting has been made on these two properties of
386 GRBs, which are observed by {\em Swift}. We also collect 114,728
solar flares from {\em RHESSI}, and find that the power law
correlation between total counts and duration of flares is
comparable with the correlation of GRBs. In order to make
comparison, we also apply this relation to superflares of solar-type
stars from {\em Kepler}.

Linear regression in log-log fields of GRBs is showed in Figure
\ref{fig:grbline}. We find the correlation of GRBs as $T_{\rm
duration} \propto {E_{\rm iso}}^{0.34\pm 0.032}$, which is
resemblant with our findings of stellar superflares as $T_{\rm
superflares} \propto {E_{\rm superflares}}^{0.39\pm 0.025}$, and
solar flares as $T_{\rm solar,duration} \propto {C_{\rm
total}}^{0.33\pm 0.001}$. The $t$-test and $F$-test show the
tendency of this correlation is genuinely credible even for
different datasets. From another aspect, our results is approximate
to the theoretical correlation $T_{\rm duration} \propto
E^{\frac{1}{3}}$, which is derived from magnetic reconnection
theorems. This comparability firmly support us to believe that
magnetic reconnection may dominate the energetic releasing process
of GRBs.

\section*{Acknowledgements}
We thank the referee for detailed and very constructive suggestions
that have allowed us to improve our manuscript. We would like to
thank H. Yu, G. Q. Zhang, J. S. Wang and H. C. Chen for suggestions.
This work is supported by the National Natural Science Foundation of
China (grant U1831207).

\begin{table*}[!p]
    \renewcommand{\arraystretch}{1}
    \addtolength{\tabcolsep}{+10pt}
    \caption{Fitting Results.}
    \label{tab:result}
    \centering
\begin{tabular}{cccc}
    \\
       \hline
       \hline
       Dataset      & GRBs                         & Solar flares                  & Stellar superflares           \\
       Satellite    & {\em Swift} & {\em RHESSI} & {\em Kepler} \\ \hline
       $k$          & 0.34 $\pm$ 0.032             & 0.33 $\pm$ 0.00093            & 0.39 $\pm$ 0.025              \\
       $b$          & -16.36 $\pm$ 1.64            & 0.95 $\pm$ 0.0045             & -12.14 $\pm$ 0.86             \\
       $\sigma_{v}$ & 0.66 $\pm$ 0.024             & 0.24 $\pm$ 0.00051            & 0.25 $\pm$ 0.013              \\
       r            & 0.47                         & 0.72                          & 0.75                          \\
       $t$          & 10.32                        & 350.82                        & 15.48                         \\
       $t_{5\%/2}$  & 1.97                         & 1.96                          & 1.97                          \\
       $F$          & 106.58                        & 123076.52                     & 239.52                        \\
       $F_{5\%}$    & 3.87                         & 3.84                          & 3.89                          \\ \hline
    \end{tabular}
\begin{flushleft}
    \textsc{Note.}\\ {In the table, $k$ and $b$ represent slope and intercept of linear regression, respectively. $\sigma_{v}$ is extra variability.
         $r$ gives correlation coefficient. $t_{5\%/2}$ and $F_{5\%}$ are upper limits of $t$ and $F$
         distribution at $5\%$ possibility. The fitting results of $t$ and $F$ are greater than $t_{5\%/2}$ and $F_{5\%}$ respectively. }\\
    \vspace{1.5em}
\end{flushleft}

\end{table*}

\end{document}